\documentclass[runningheads]{llncs}
\PassOptionsToPackage{table}{xcolor} 

\usepackage{hyperref}

\usepackage{tikz}
\usepackage{xcolor} 

\usepackage[T1]{fontenc}
\usepackage{graphicx}
\usepackage{booktabs}
\usepackage{multicol}
\usepackage{multirow}
\usepackage{xspace}
\usepackage{adjustbox}
\usepackage{subcaption}

\newcommand{\partitle}[1]{\vspace{2mm}\noindent\textbf{#1}}

\newcommand{\IGNORE}[1]{}

\newcommand{\model}{\texttt{JudgeBlender}\xspace}
\newcommand{\prompt}{\texttt{PromptBlender}\xspace}
\newcommand{\llm}{\texttt{LLMBlender}\xspace}

\begin{document}
\title{JudgeBlender: Ensembling Judgments for Automatic Relevance Assessment}
\titlerunning{JudgeBlender: Ensembling Judgments for Automatic Relevance Assessment}
%
\author{Hossein A.~Rahmani\inst{1} \and Emine Yilmaz\inst{1,2,3} \and Nick Craswell\inst{4} \and Bhaskar Mitra\inst{5}}
\authorrunning{H.~A.~Rahmani et al.}
\institute{University College London, London, UK \and
The Alan Turing Institute, London, UK \and
Amazon, London, UK \and
Microsoft, Seattle, USA \and
Microsoft Research, Montréal, Canada \\
\email{hossein.rahmani.22@ucl.ac.uk}}
\maketitle              
\begin{abstract}
The effective training and evaluation of retrieval systems require a substantial amount of relevance judgments, which are traditionally collected from human assessors -- a process that is both costly and time-consuming. Large Language Models (LLMs) have shown promise in generating relevance labels for search tasks, offering a potential alternative to manual assessments. Current approaches often rely on a single LLM, such as GPT-4, which, despite being effective, are expensive and prone to intra-model biases that can favour systems leveraging similar models. In this work, we introduce \textbf{\model}, a framework that employs smaller, open-source models to provide relevance judgments by combining evaluations across multiple LLMs (\llm) or multiple prompts (\prompt). By leveraging the LLMJudge benchmark \cite{rahmani2024llmjudge}, we compare JudgeBlender with state-of-the-art methods and the top performers in the LLMJudge challenge. Our results show that JudgeBlender achieves competitive performance, demonstrating that very large models are often unnecessary for reliable relevance assessments.
\end{abstract}
\section{Introduction}
\label{sec:introduction}

In Information Retrieval (IR), large-scale datasets that capture the relevance of documents to users' queries are critical for training and evaluating retrieval systems. Traditionally, relevance judgments have been obtained either manually, through human assessors, or via heuristic-based methods. While effective, these approaches face limitations such as scalability issues, human error, and subjective biases that can skew evaluations. With the rapid advancements in Large Language Models (LLMs), there is a growing opportunity to automate the relevance judgment process by leveraging their capabilities to comprehend and reason over large volumes of documents and passages.

LLMs have demonstrated impressive performance across various natural language processing tasks, including text classification, summarisation, and complex reasoning. However, when applied to the task of generating relevance judgments in IR, individual models may exhibit inherent biases, performance inconsistencies across domains, and susceptibility to overfitting specific linguistic patterns. These challenges make it difficult to rely solely on a single LLM for reliable relevance scoring across diverse datasets and query types.

To address these issues, we propose a novel framework, \texttt{\model}, that employs an ensemble of LLMs to generate more robust and accurate relevance judgments. Instead of depending on a single model as a ``judge,'' our approach introduces a panel of diverse evaluators, each contributing unique perspectives to the relevance evaluation process. By aggregating their outputs, we aim to achieve a more balanced and comprehensive assessment. This ``jury'' of models produces multiple relevance scores for each query-document pair, which are subsequently aggregated using various ensemble strategies -- such as averaging, weighted voting, and advanced statistical methods -- to yield a final, more reliable relevance score.

The key advantage of this approach lies in its ability to leverage the strengths of different LLMs while minimising their individual weaknesses. For instance, some models may excel at identifying semantic similarities in short texts, whereas others are better suited to processing longer, more complex documents. By ensembling models, we harness their complementary strengths, resulting in a more consistent and accurate determination of relevance.

Our contributions in this work are threefold. First, we develop a methodology for ensembling LLMs tailored to the task of generating relevance judgments. Second, we design and implement multiple aggregator functions to combine individual model outputs in ways that optimise the final relevance score. Third, we conduct extensive experiments on the LLMJudge challenge dataset \cite{rahmani2024llmjudge}, demonstrating that our ensemble-based approach outperforms individual LLMs by achieving higher precision and consistency in relevance assessments.
\section{Related Work}
\label{sec:relatedwork}

The considerable time and effort required to prepare manual tests for large-scale evaluations have long driven researchers to explore automated methods for collecting relevance judgments \cite{faggioli2023perspectives}. These systems aim to provide an efficient and scalable alternative that achieves comparable alignment with manual assessments, offering a practical solution for relevance evaluation in information retrieval systems. With the rapid progress of Large Language Models (LLMs) across diverse tasks, researchers have increasingly examined their potential to automate the relevance judgment process \cite{faggioli2023perspectives}. However, challenges such as bias, inconsistency, and limited generalisation persist when relying on a single model for relevance assessments.

\begin{figure}[t]
    \centering
    \includegraphics[width=\linewidth]{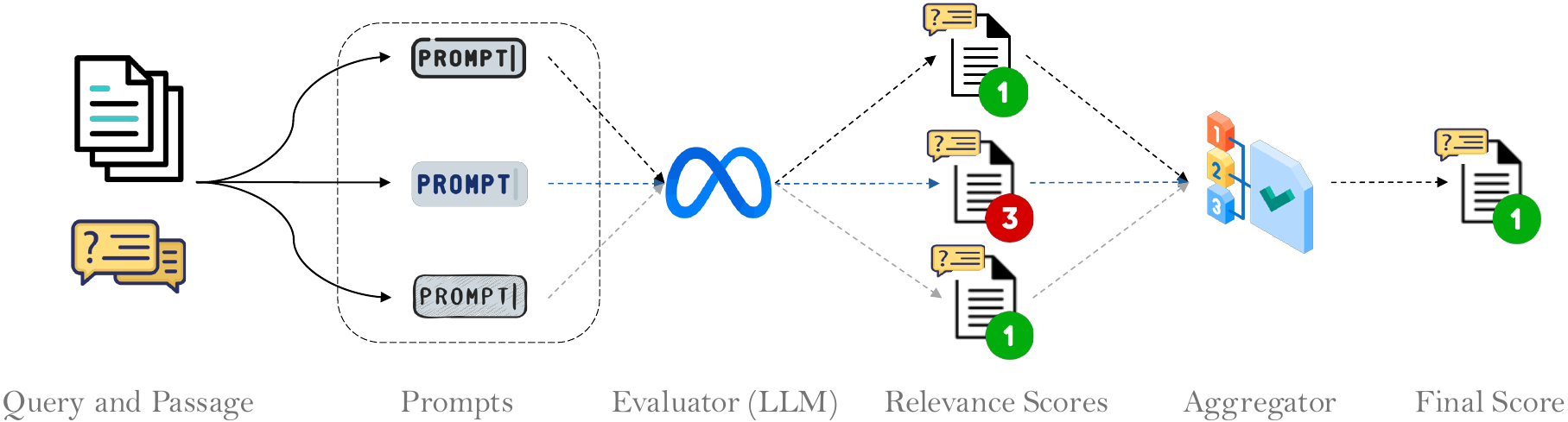}
    \caption{The \prompt evaluation uses an LLM to grade how relevant a passage is to a query. The evaluation is done by multiple prompting to an LLM then aggregating the scores based on an aggregation function (Section \ref{sec:aggregation-function}).}
    \label{fig:promptblender}
\end{figure}

Ensemble methods have been extensively adopted in machine learning to enhance performance by leveraging the strengths of multiple models \cite{dietterich2000ensemble,xu2007adarank,polikar2012ensemble,sagi2018ensemble}. By incorporating the diversity of individual models, ensemble approaches improve the reliability of predictions, particularly in noisy or uncertain environments \cite{breiman1996bagging,jahrer2010combining,aniol2019ensemble}. Typically, these methods combine models through weighted contributions or by aggregating their diverse outputs, demonstrating the effectiveness of incorporating multiple perspectives to achieve robust results.

In recent years, ensemble methods have been applied to LLMs for tasks such as summarisation \cite{ravaut2022towards,llm-blender-2023,verga2024replacing}, translation \cite{llm-blender-2023}, and question answering \cite{izacard2020leveraging,verga2024replacing}. For instance, Jiang et al.~\cite{llm-blender-2023} demonstrated a method that combines outputs from multiple LLMs using pairwise comparisons to refine subtle differences between outputs, producing a more accurate final result. Similarly, Verga et al.~\cite{verga2024replacing} introduced a panel of LLMs as evaluators to assess free-form generation outputs, illustrating the potential of ensembling for improving LLM-based evaluations.

Recent studies on automated relevance judgments using LLMs have primarily focused on different prompting techniques, including zero-shot, one-shot, and few-shot learning \cite{faggioli2023perspectives,thomas2024large,abbasiantaeb2024can,upadhyay2024umbrela,upadhyay2024llms}. Researchers have found that detailed instructions, defined roles, and multiple evaluators enhance the alignment between LLM-generated assessments and human evaluations. While earlier studies generated relevance labels using basic zero-shot prompts \cite{faggioli2023perspectives}, more recent work has shown that explicit instructions for analysing query intent and trustworthiness improve output quality \cite{thomas2024large}. For example, Upadhyay et al.~\cite{upadhyay2024umbrela} proposed UMBRELA, an open-source framework that applied structured prompts from Thomas et al.~\cite{thomas2024large} using GPT-4o. In another study, Farzi and Dietz~\cite{farzi2024pencils} introduced the RUBRIC metric, which evaluates passages based on query-specific questions. Despite the progress in using LLMs for automated relevance judgments, many of these studies rely on commercial models (e.g., ChatGPT, GPT-3.5/4), which face limitations such as high costs, reproducibility challenges, nondeterministic outputs, and potential risks of data leakage.

To the best of our knowledge, no prior work has investigated the use of LLM ensembles for automated relevance judgments in IR. While existing research has explored ensembling for tasks like summarisation or classification, our work uniquely focuses on combining outputs from multiple LLMs for relevance judgments. Specifically, we integrate diverse prompt strategies (\prompt) and model scores (\llm) to provide a broader range of perspectives and mitigate the inherent biases of individual models. By extending ensembling techniques to the domain of relevance assessment, we address the precision and contextual understanding necessary for high-quality relevance annotations.
\section{JudgeBlender}
\label{sec:method}
\model is based on using an ensemble of models such that the final output is obtained by aggregating the outputs of these models. \model has two variants: \prompt and \llm.

\partitle{PromptBlender.} In this approach, a single language model is used with a variety of distinct prompts to assess the relevance of a query to a passage. The key idea behind \prompt is to leverage prompt diversity, ensuring that different perspectives on the same task are captured. Each prompt is designed to elicit different reasoning pathways or interpretations of relevance from the model. The outputs from these various prompts are aggregated, typically by averaging the relevance scores or using a voting function to select the most frequent or consensus relevance level. This method allows for a nuanced and multi-faceted evaluation of relevance while relying on the underlying capabilities of a single model.

\begin{figure*}[t]
    \centering
    \includegraphics[width=\linewidth]{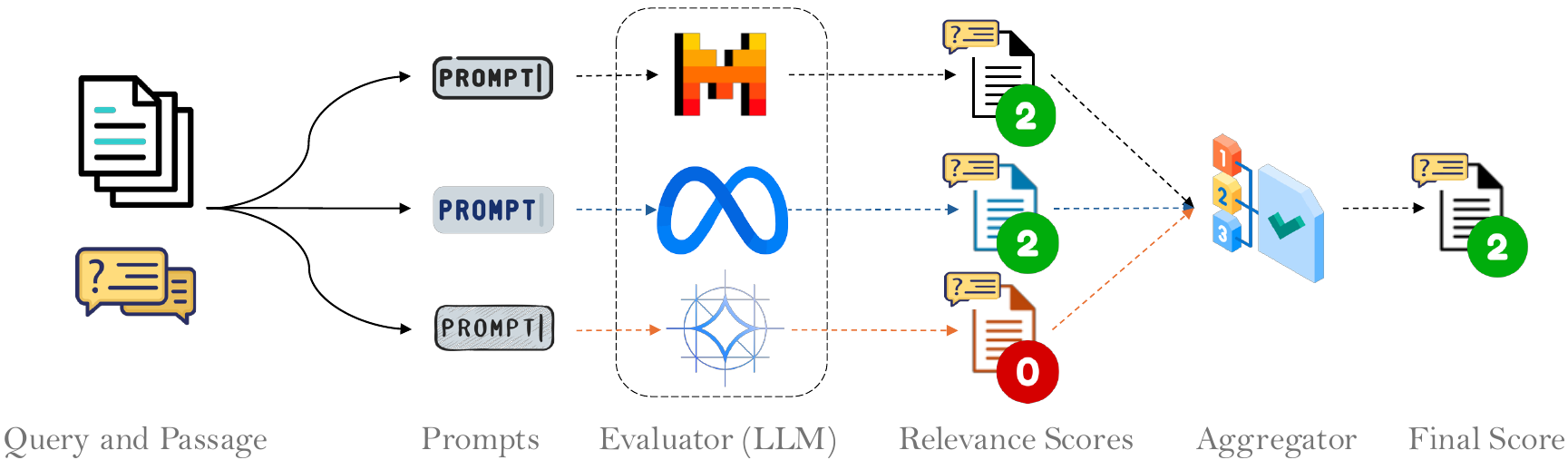}
    \caption{The \llm evaluation uses multiple LLMs to grade how relevant a passage is to a query. The evaluation is done by prompting several LLMs, then aggregating the scores based on an aggregation function (Section \ref{sec:aggregation-function}).}
    \label{fig:llmblender}
\end{figure*}

\partitle{LLMBlender.} \llm extends the ensembling strategy by incorporating multiple language models, each tasked with evaluating the query-to-passage relevance based on their own prompt. In this variant, instead of using one model with diverse prompts, multiple models are employed, each potentially offering different strengths in understanding and contextualising information. Each language model is prompted with a carefully designed instruction to elicit relevance judgments. The evaluations from these models are aggregated similarly to \prompt, either by averaging the scores or through another fusion mechanism. This allows for a more diverse set of judgments, harnessing the complementary strengths of various language models to arrive at a more robust and reliable final relevance score.

To calculate the final relevance score, the individual scores are pooled together through an aggregator function such that the final score $= f (j \in P : j(a))$ where $P \in \{\prompt, \llm\}$ is a panel composed of individual judges $j$ and $f$ is an aggregator function.
\section{Experimental Setup}
\label{sec:experiment}

\begin{table}[t]
    \centering
    \caption{Statistics of LLMJudge challenge dataset}
    \begin{adjustbox}{max width=\textwidth}
    \label{tbl:llmjudge-dataset}
        \begin{tabular}{lccc|cccc}
            \toprule
            \textbf{} & \textbf{\#queries} & \textbf{\#passage} & \textbf{\#qrels} & \textbf{irrelevant} & \textbf{related} & \textbf{high.~rel} & \textbf{perfect.~rel} \\
            \midrule
             Dev  & 25   & 7,224 & 7,263 & 4,538 & 1,403 & 625 & 697 \\
             Test & 25   & 4,414 & 4,423 & 2,005 & 1,233 & 808 & 377 \\
            \bottomrule
        \end{tabular}
        \end{adjustbox}
\end{table}

\subsection{Dataset}
In our experiments, we used the LLMJudge challenge dataset \cite{rahmani2024llmjudge,rahmani2024report,rahmani2024llm4eval}, which is based on the TREC DL 2023 \cite{craswell2024overview,rahmani2024synthetic} passage ranking task with human judgments (also known as \textit{qrels}). A TREC DL judgment includes the query, passage, and a relevance label assigned to the passage by human experts. Relevance scores are on a four-point scale: perfectly relevant (3), highly relevant (2), related (1), and irrelevant (0). Table \ref{tbl:llmjudge-dataset} shows the detailed statistics of the LLMJudge challenge dataset.

\subsection{Aggregator Function}
\label{sec:aggregation-function}
We explore two different aggregation functions to combine scores from multiple judges: (i) majority voting (MV) and (ii) average voting (AV). For majority voting, in the case of a tie, we apply four strategies to resolve the conflict: (1) selecting a random score (Rnd), (2) choosing the maximum score (Max), (3) choosing the minimum score (Min), or (4) taking the average (Avg) of the tied scores. This allows us to examine how different tie-breaking approaches impact the final relevance judgment. For average voting, we directly compute the average of the judges' scores as the final relevance score. Future work could explore using another LLM to break ties or make the final decision.

\subsection{Models and Prompt Families}
We use open-source, small language models for both variants of \model. For the \prompt method, we employ \texttt{Meta-Llama-3-8B} as our base model and adopt three distinct prompting strategies to generate the pool of judgments:
(1) the prompt proposed by Thomas et al.~\cite{thomas2024large},
(2) a prompt that breaks down the concept of ``relevance’’ into multiple criteria inspired by \cite{farzi2024best}, and
(3) a two-step prompt that first asks for binary relevance judgment, followed by generating final scores for relevant passages.
More details on \prompt can be found in Section \ref{sec:baselines}, and the prompts are available on our GitHub\footnote{\url{https://github.com/rahmanidashti/JudgeBlender}}.

For the \llm method, judgments are derived from three distinct models belonging to different model families. Specifically, we consider \texttt{Mistral-7B}, \texttt{Gemma-7B}, and \texttt{Llama-3-8B} in our experiments. As illustrated in Figure \ref{fig:llmblender}, each model is prompted using a different variant: (1) For \texttt{Mistral-7B}, we use the prompt that decomposes ``relevance’’ into multiple criteria. For \texttt{Gemma-7B}, we employ the prompt from Thomas et al.~\cite{thomas2024large}, and (3) for \texttt{Llama-3-8B}, we apply the two-step prompt, which first asks for binary relevance judgment and then generates final scores for relevant passages.

\subsection{Evaluation Measurement}
Following recent studies \cite{faggioli2023perspectives,farzi2024pencils,thomas2024large}, we evaluate the proposed methods and baseline approaches using two levels of measurement: (i) label correlation and (ii) system ranking correlation.

\subsubsection{Correlation to Human Judgments.}
We measure label correlation using Cohen's $\kappa$ and Krippendorff's $\alpha$.

\begin{itemize}
    \item \textbf{Cohen’s $\kappa$} quantifies inter-rater reliability by assessing agreement between raters while accounting for chance. A value of $\kappa > 0.8$ indicates strong agreement, while $\kappa > 0.6$ indicates moderate agreement.
    \item \textbf{Krippendorff’s $\alpha$} is a more flexible measure, accommodating multiple raters and handling various data types, including missing data. It generalizes Cohen’s $\kappa$ and similarly interprets $\alpha > 0.8$ as strong agreement and $\alpha > 0.6$ as moderate agreement.
\end{itemize}

Both metrics are used to evaluate the agreement between human labels and the relevance scores generated by our LLM-based methods.

\subsubsection{System Ranking Correlation.}
We measure system ranking correlation using Spearman's $\rho$ and Kendall's $\tau$. Both metrics range from $-1.0$ (complete disagreement) to $1.0$ (perfect agreement), with $0$ indicating no correlation in the system rankings.

\begin{itemize}
    \item \textbf{Kendall's $\tau$} evaluates the similarity between two sets of rankings by comparing the number of concordant and discordant pairs.
    \item \textbf{Spearman's $\rho$} assesses the rank correlation by measuring the monotonic relationship between two rankings.
\end{itemize}

These metrics allow us to analyze how well our methods preserve system rankings relative to human-based rankings.

\subsection{Comparison Methods}
\label{sec:baselines}
We compare the variations of \model including \prompt and \llm as well as recent state-of-the-art automatic relevance judgment baselines:

\begin{itemize}
    \item \textbf{Faggioli et al}~\cite{faggioli2023perspectives}: We used the updated instructions from Figure 2 \cite{faggioli2023perspectives} by adding the definition of the relevance scores from TREC DL 2023 \cite{craswell2024overview} and asking LLMs to generate the relevance score instead of binary classification.
    
    \item \textbf{Thomas et al}~\cite{thomas2024large}: We used the general relevance direct grading prompt which includes the role feature (see Figure 2 \cite{thomas2024large}).
    
    \item \textbf{MultiCriteria}~\cite{farzi2024best}: This is the best-performing method from the LLMJudge challenge \cite{rahmani2024llmjudge}. This method evaluates the relevance of a passage to a query by breaking down the concept of ``relevance'' into four criteria: Exactness, Coverage, Topicality, and Contextual Fit. Each criterion is individually assessed through prompting an LLM, and the resulting scores are prompted for the final relevance judgment score.
    
    \item \textbf{Rubric (Question)}~\cite{farzi2024pencils}: This method assesses relevance by evaluating how well a passage answers about 10 open-ended questions, assigning grades to each. The final relevance label is determined using a heuristic mapping of grades to relevance scores.

    \item \textbf{GenRE}~\cite{meng2024query}: This method fine-tuned Llama model under two different settings for 5 epoch. \textbf{GenRE-dev}: Fine-tune Llama-3-8B on the dev set of LLMJudge dataset \cite{rahmani2024llmjudge}. \textbf{GenRE-trec}: Fine-tune Llama-3-8B on the qrels of TREC-DL 2019, 2020, and 2021, as well as the dev set of the LLMJudge dataset \cite{rahmani2024llmjudge}.

    \item \textbf{SunMulti}: This method applies approach proposed by Sun et al.~\cite{sun2023chatgpt} to give a binary relevance judgment, and then generates relevance scores (1-3) only for passages marked as relevant.

    \item \textbf{RelExp}: This method includes reasoning for the relevance judgment within the prompt prposed by Thomas et al.~\cite{thomas2024large}, asking the LLM to explain its assessment while rating the passage based on specific criteria.

    \item \textbf{PromptBlender}: \prompt is a variant of \model that uses same LLM with different prompting techniques. We use three different prompting strategies with \texttt{Llama-3-8B} in our experiments. We use (1) prompt proposed by Thomas et al.~\cite{thomas2024large} for \prompt{}$_1$, (2) prompt that used by \textbf{MultiCriteria} for \prompt{}$_2$, and (3) prompt proposed by Sun et al.~\cite{sun2023chatgpt} for \prompt{}$_3$. Then, we aggregate their outputs based on aggregation functions defined in Section \ref{sec:aggregation-function}.
    
    \item \textbf{LLMBlender}: \llm is a variant of \model that uses different LLMs with different prompting techniques. We consider (1) \texttt{Mistral-7B} with the prompt used by by \textbf{MultiCriteria} method, (2) \texttt{Gemma-7B} with prompt proposed by Thomas et al.~\cite{thomas2024large}, and \texttt{Llama-3-8B} model with prompt proposed by Sun et al.~\cite{sun2023chatgpt}. Similarly to \prompt, the final score is computed by aggregating the scores generated by each model.
\end{itemize}
\section{Results}
\label{sec:results}

\begin{table*}
    \centering
    \caption{Judgment and system ranking correlation of \model methods compared to \model variation, direct LLM relevance label prompts, fine-tuned methods, and methods based on GPT-4o. $\kappa$: Cohen's Kappa, $\alpha$: Krippendorff's alpha, $\tau$: Kendall's Tau, $\rho$: Spearman's rank correlation. Best results per column denoted in \colorbox{lime}{\textbf{green-bold}}, best across baselines methods denoted in \colorbox{cyan!20}{cyan}, and best per aggregator function for each variation of \model is denoted in \colorbox{yellow!50}{yellow}.}
    \label{tbl:result}
    \begin{adjustbox}{max width=\textwidth}
        \begin{tabular}{lllccccc}
            \toprule
            \textbf{Method} & \textbf{Model} & $\kappa$ & $\alpha$ & \multicolumn{2}{c}{$\tau$} & \multicolumn{2}{c}{$\rho$}\\
            &&&& \textbf{NDCG@10} & \textbf{MAP} & \textbf{NDCG@10} & \textbf{MAP}\\
            \midrule
            \multicolumn{8}{c}{\textbf{Baselines}} \\
            \midrule
             Faggioli et al.~\cite{faggioli2023perspectives} & GPT-35-turbo & 0.0754 & 0.2808 & 0.9181 & 0.9054 & 0.9863 & 0.9798\\
             Faggioli et al.~\cite{faggioli2023perspectives} & GPT-35-turbo-16k & 0.1327 & 0.3672 & 0.8966 & 0.8796 & 0.979 & 0.9726\\
             Faggioli et al.~\cite{faggioli2023perspectives} & GPT-4-32k & 0.211 & 0.4642 & 0.9052 & 0.8796 & 0.981 & 0.9698\\ 
             \midrule
             Thomas et al.~\cite{thomas2024large} & GPT-35-turbo & 0.1236 & 0.3207 & 0.8664 & 0.8968 & 0.9689 & 0.9798\\
             Thomas et al.~\cite{thomas2024large} & GPT-35-turbo-16k & 0.1723 & 0.3853 & 0.8793 & 0.9054 & 0.975 & 0.9827\\
             Thomas et al.~\cite{thomas2024large} & GPT-4-32k & 0.2293 & \cellcolor{cyan!20}0.4877 & 0.9181 & 0.9011 & 0.9867 & 0.9778\\
            \midrule
             MultiCriteria \cite{farzi2024best} & Llama-3-8B-Instruct & 0.1829 & 0.2888 & \cellcolor{cyan!20}0.9483 & 0.9140 & \cellcolor{cyan!20}0.9919 & 0.9794 \\
             Rubric (Question)~\cite{farzi2024pencils} & GPT-3.5/FlanT5-large & 0.0779 & 0.1036 & 0.8276 & 0.8839 & 0.9544 & 0.9714 \\
            \midrule
             \multicolumn{8}{c}{\textbf{Fine-tuned Methods}} \\
             \midrule
             GenRE-dev~\cite{meng2024query} & Llama-3-8B-Instruct & 0.1823 & 0.4069 & 0.9042 & \cellcolor{lime}\textbf{0.9312} & 0.9826 & \cellcolor{lime}\textbf{0.9879} \\	
             GenRE-trec~\cite{meng2024query} & Llama-3-8B-Instruct & 0.1471 & 0.1623 & 0.8568 & 0.9011 & 0.9608 & 0.9806 \\
             \midrule
             \multicolumn{8}{c}{\textbf{Methods based on GPT-4o}} \\
             \midrule
             SunMulti~\cite{sun2023chatgpt} & GPT-4o & 0.2388 & 0.4108 & 0.8966 & 0.8968 & 0.9798 & 0.977 \\	
             RelExp & GPT-4o & \cellcolor{cyan!20}0.2519 & 0.4701 & 0.9009 & 0.9140 & 0.9819 & 0.9847 \\
             \midrule
             \multicolumn{8}{c}{\textbf{PromptBlender}} \\
             \midrule
             \prompt{}$_1$ & \multirow{3}{*}{Llama-3-8B-Instruct} & 0.0465 & 0.1192 & 0.9042 & 0.8882 & 0.9822 & 0.9762\\
             \prompt{}$_2$ & & 0.1741 & 0.3579 & 0.9128 & 0.8827 & 0.9838 & 0.9745\\
             \prompt{}$_3$ && 0.2374 & 0.4482 & 0.9136 & 0.8764 & 0.9626 & 0.9649\\
             \midrule
             \multicolumn{8}{l}{\prompt} \\
             + MV(Avg.) & & 0.2398 & 0.4769 & \cellcolor{yellow!50}0.9526 & 0.8968 & \cellcolor{yellow!50}0.9919 & 0.9762 \\
             + MV(Rnd.) & & \cellcolor{yellow!50}0.2436 & 0.4747 & 0.931 & 0.8925 & 0.9875 & 0.9790 \\
             + MV(Max.) & Llama-3-8B-Instruct & 0.2219 & 0.4527 & 0.9085 & 0.8968 & 0.9838 & 0.9762 \\
             + MV(Min.) &  & 0.2023 & 0.4052 & 0.9085 & 0.8839 & 0.9813 & 0.9758 \\
             + AV &  & 0.2379 & \cellcolor{lime}\textbf{0.4887} & 0.9224 & \cellcolor{yellow!50}0.9054 & 0.9863 & \cellcolor{yellow!50}0.9798 \\
             \midrule
             \multicolumn{8}{c}{\textbf{LLMBlender}} \\
             \midrule
             \llm{}$_1$ & Mistral 7B & 0.0832 & 0.1110 & 0.9267 & 0.8968 & 0.9867 & 0.977 \\
             \llm{}$_2$ & Gemma 7B & 0.1880 & 0.3821 & 0.9440 & 0.9069 & 0.9907 & 0.9542\\
             \llm{}$_3$ & Llama-3-8B & 0.2454 & 0.4673 & 0.9353 & 0.8968 & 0.9883 & 0.9786 \\
             \midrule	
             \multicolumn{8}{l}{\llm} \\
             + MV(Avg.) && 0.2553 & 0.4784 & \cellcolor{lime}\textbf{0.9612} & 0.9011 & \cellcolor{lime}\textbf{0.9940} & 0.9806 \\
             + MV(Rnd.) & Mistral 7B & \cellcolor{lime}\textbf{0.2619} & 0.4772 & 0.9569 & 0.9011 & \cellcolor{lime}\textbf{0.9940} & 0.9810\\
             + MV(Max.) & Gemma 7B & 0.2543 & 0.4620 & 0.9397 & 0.9011 & 0.9899 & 0.9806 \\
             + MV(Min.) & Llama-3-8B & 0.2600 & 0.4709 & 0.9483 & 0.9011 & 0.9923 & 0.9810 \\
             + AV && 0.2502 & \cellcolor{yellow!50}0.4832 & 0.9569 & \cellcolor{yellow!50}0.9054 & 0.9935 & \cellcolor{yellow!50}0.9815 \\
            \bottomrule
        \end{tabular}
        \end{adjustbox}
\end{table*}

\subsection{Correlation to Human Judgments}
Table \ref{tbl:result} presents the correlation between scores produced by different evaluator methods and human judgments, measured using Cohen's $\kappa$ and Krippendorff's $\alpha$. Overall, both variants of \model achieve the strongest correlation across both metrics. In contrast, methods based on GPT-4 and fine-tuned approaches show weaker performance, particularly when evaluated using Cohen’s $\kappa$.

\subsection{System Ranking Correlation}
Table \ref{tbl:result} also shows how system rankings produced by different evaluation methods correlate with human judgments on TREC DL 2023 submissions. We report Kendall's $\tau$ and Spearman's rank correlation ($\rho$) between the system rankings generated by each method and those based on human judgments.

We observe that \llm achieves the highest correlation with human rankings when NDCG@10 is used as the evaluation metric. However, the results for MAP reveal slight differences: the fine-tuned method on the development set of LLMJudge and GPT-4o, when prompted to provide explanations during judgment, demonstrate better performance.

\begin{table*}[t]
    \centering
    \caption{Judgments inter-annotator agreement on LLMJudge challenge dataset. Comparing best variants of \prompt and \llm to best baseline methods in terms of Cohen's $\kappa$ and NDCG@10 (i.e., MultiCriteria and RelExp). Percentage shows the accurate percentage judgment for each judgment level.}
    \label{tbl:agreement}
    \begin{adjustbox}{max width=\textwidth}
        \begin{tabular}{lccccccccccc}
            \toprule
            \multirow{2}{*}{\textbf{TREC}} && \multicolumn{4}{c}{\textbf{MultiCriteria}} & \multirow{2}{*}{\textbf{\%}} & \multicolumn{4}{c}{\textbf{RelExp}} & \multirow{2}{*}{\textbf{\%}} \\
            \cmidrule{3-6}\cmidrule{8-11}
            && Irrelevant & Related & High.~rel. & Perfect.~rel. && Irrelevant & Related & High.~rel. & Perfect.~rel. & \\
            \midrule
             Perfect.~rel. && 10 & 26 & 243 & 98 & 25.99\% & 61 & 126 & 68 & 122 & 32.36\% \\
             High.~rel. && 43 & 72 & 596 & 97 & 73.76\% & 206 & 307 & 186 & 109 & 23.01\% \\
             Related && 191 & 244 & 682 & 116 & 19.78\% & 618 & 417 & 136 & 62 & 33.81\% \\
             Irrelevant && 783 & 409 & 692 & 121 & 39.05\% & 1550 & 360 & 66 & 29 & 77.30\%\\
             \midrule
             && \multicolumn{4}{c}{\textbf{PromptBlender - MV(Avg.)}} && \multicolumn{4}{c}{\textbf{LLMBlender - MV(Avg.)}} \\
            \cmidrule{3-6}\cmidrule{8-11}
            && Irrelevant & Related & High.~rel. & Perfect.~rel. && Irrelevant & Related & High.~rel. & Perfect.~rel.\\
            \midrule
            Perfect.~rel. && 45 & 24 & 271 & 37 & 9.81\% & 25 & 47 & 204 & 101 & 26.79\% \\
            High.~rel. && 150 & 112 & 510 & 36 & 63.11\% & 75 & 179 & 479 & 75 & 59.28\%\\
            Related && 583 & 157 & 456 & 37 & 12.73\% & 393 &351 &410 & 79 & 28.46\% \\
            Irrelevant && 1454 & 162 & 375 & 14 & 72.51\% & 1189 & 429 & 340 & 47 & 59.75\%\\
            \midrule
            \bottomrule
        \end{tabular}
        \end{adjustbox}
\end{table*}

\subsection{Inter-Judge Agreement Analysis}
We analyze the agreement between the manual TREC DL 2023 judgments (LLMJudge challenge dataset) and the predicted relevance labels, as presented in Tables \ref{tbl:agreement} and \ref{tbl:binary-agreement}. The inter-annotator agreement metric measures the percentage of correctly predicted judgments at each relevance level, providing insights into how closely the model's predictions align with human judgments. For binary relevance judgments, following the TREC DL 2023 track guidelines \cite{craswell2024overview}, levels 0 and 1 are classified as irrelevant, while levels 2 and 3 are considered relevant.

In our analysis, the \textit{MultiCriteria} method achieves the highest agreement at the very relevant level, particularly for highly and perfectly relevant passages, highlighting its strength in identifying strong relevance. Conversely, the \textit{RelExp} method demonstrates higher agreement with irrelevant passages, suggesting it is more effective at detecting non-relevant content.

However, the best-performing variant of \model, \llm{} - MV(Avg.), demonstrates consistent and strong agreement across all four relevance levels. This balanced performance becomes even more evident when the judgment scale is collapsed into a binary format, where levels 0 and 1 represent irrelevant content and levels 2 and 3 indicate relevant content. In this binary setting, \llm{} - MV(Avg.) maintains reliable agreement, underscoring its versatility and accuracy in capturing relevance across the entire judgment scale.

\begin{table*}[t]
    \centering
    \caption{Judgments inter-annotator agreement at binary level on LLMJudge challenge dataset. Comparing best variants of \prompt and \llm to best baseline methods in terms of Cohen's $\kappa$ and NDCG@10 (i.e., MultiCriteria and RelExp). Percentage shows the accurate percentage judgement for each judgment level.}
    \label{tbl:binary-agreement}
    \begin{adjustbox}{max width=\textwidth}
        \begin{tabular}{lccccccc}
            \toprule
            \multirow{2}{*}{\textbf{TREC}} && \multicolumn{2}{c}{\textbf{MultiCriteria}} & \multirow{2}{*}{\textbf{\%}} & \multicolumn{2}{c}{\textbf{RelExp}} & \multirow{2}{*}{\textbf{\%}} \\
            \cmidrule{3-4}\cmidrule{6-7}
            && Irrelevant & Relevant && Irrelevant & Relevant & \\
            \midrule
             Relevant && 151 & 1,034 & 87.25\% & 700 & 485 & 40.92\% \\
             Irrelevant && 1,627 & 1,611 & 49.75\% & 2,945 & 293 & 90.95\% \\
             \midrule
            \multirow{2}{*}{\textbf{TREC}} && \multicolumn{2}{c}{\textbf{MultiCriteria}} & \multirow{2}{*}{\textbf{\%}} & \multicolumn{2}{c}{\textbf{RelExp}} & \multirow{2}{*}{\textbf{\%}} \\
            \cmidrule{3-4}\cmidrule{6-7}
            && Irrelevant & Relevant && Irrelevant & Relevant & \\
            \midrule
            Relevant && 331 & 854 & 72.06\% & 326 & 859 & 72.48\% \\
            Irrelevant && 2,356 & 882 & 72.76\% & 2,362 & 876 & 72.94\% \\
            \midrule
            \bottomrule
        \end{tabular}
        \end{adjustbox}
\end{table*}

\subsection{Bias in System Evaluation}
Here, we analyse the bias that evaluating on LLM-generated relevance judgment may exhibit towards systems that are based on a similar language model to the one that was used in the relevange judgment process. To do this, similar to Rahmani et al.~\cite{rahmani2024synthetic}, we categorised the systems submitted to TREC DL 2023 based on the approach they use (i.e., language models used in their ranking or retrieval pipeline) using the metadata file released as part of SynDL\footnote{\url{https://rahmanidashti.github.io/SynDL/}} resource \cite{rahmani2024syndl}. This results in four different system categories: systems based on \textbf{GPT}, \textbf{T5}, \textbf{GPT + T5} (i.e., a combination of GPT and T5), and \textbf{others} (i.e., traditional methods such as BM25, or any model that does not use either GPT or T5).

\begin{figure}[t]
    \centering
    \begin{subfigure}[b]{0.45\textwidth}
        \includegraphics[width=\textwidth]{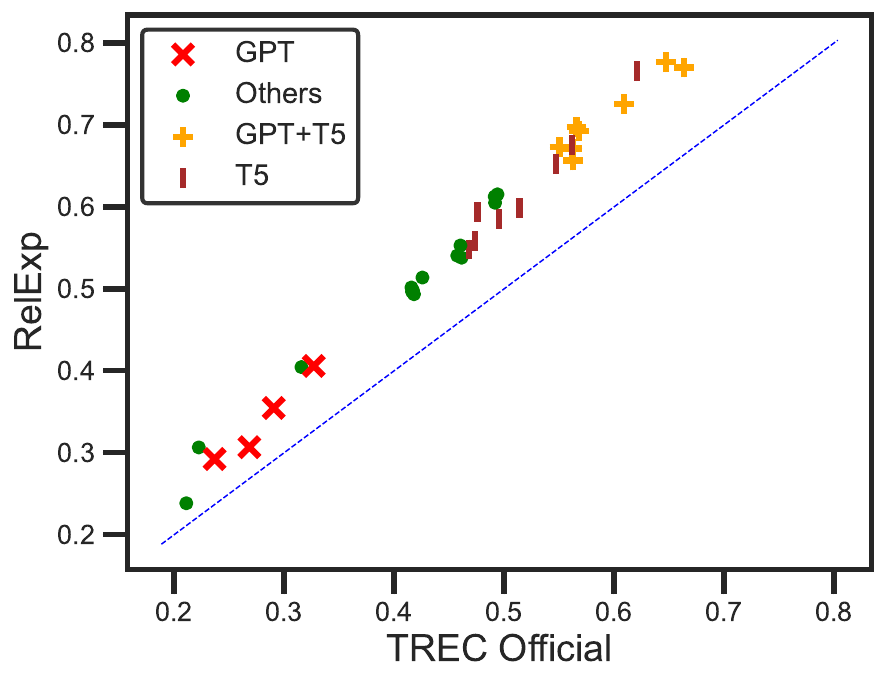}
        \caption{}
        \label{fig:relexp}
    \end{subfigure}
    \hfill
    \begin{subfigure}[b]{0.45\textwidth}
        \includegraphics[width=\textwidth]{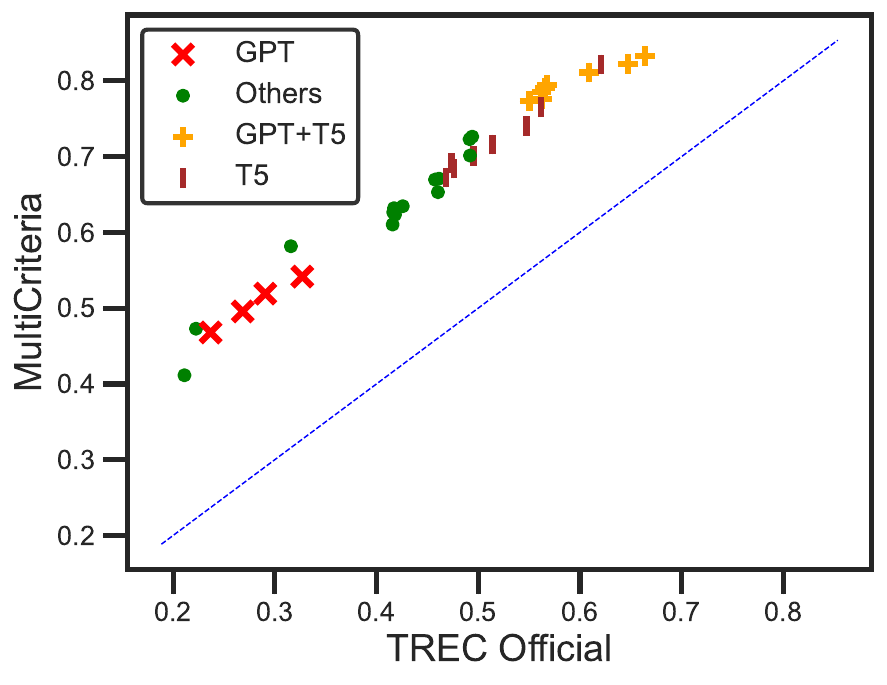}
        \caption{}
        \label{fig:multicriteria}
    \end{subfigure}
        \hfill
    \begin{subfigure}[b]{0.45\textwidth}
        \includegraphics[width=\textwidth]{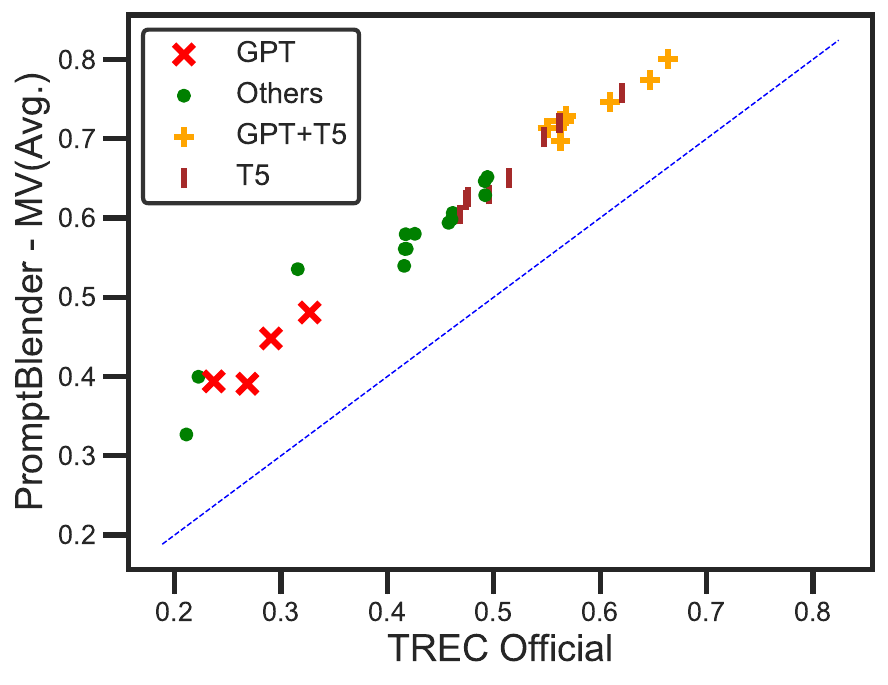}
        \caption{}
        \label{fig:pblender}
    \end{subfigure}
        \hfill
    \begin{subfigure}[b]{0.45\textwidth}
        \includegraphics[width=\textwidth]{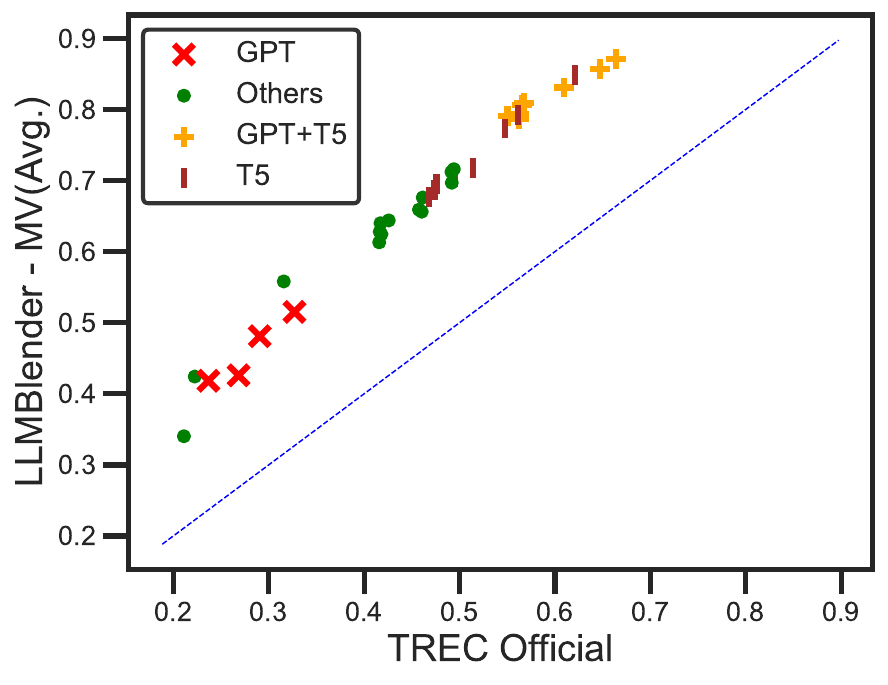}
        \caption{}
        \label{fig:lblender}
    \end{subfigure}
    \caption{Scatter plots of the effectiveness of TREC Deep Learning track 2023 runs according to the TREC official human judgments and (a) RelExp, (b) MultiCriteria, (c) PromptBlender - MV(Avg.), and (b) LLMBlender - MV(Avg.) evaluated using NDCG@10.}
    \label{fig:bias-analysis}
\end{figure}

Figure \ref{fig:bias-analysis} shows the effectiveness of different methods in predicting relevance judgments, compared to official human judgments (x-axis) across four methods: \textit{RelExp}, \textit{MultiCriteria}, \textit{\prompt{} - MV(Avg.)}, and \textit{\llm{} - MV(Avg.)}, evaluated using NDCG@10. RelExp (Figure \ref{fig:relexp}) and MultiCriteria (Figure \ref{fig:multicriteria}) show distinct patterns. The \textit{RelExp} method tends to overestimate for top-performing methods, GPT+T5 systems and T5-based systems. Conversely, \textit{MultiCriteria} exhibits a noticeable overestimation of relevance for lower-performed models, comparing GPT-based systems (red crosses) in Figures \ref{fig:relexp} and \ref{fig:multicriteria}. The best-performing methods, \prompt{} - MV(Avg.) (Figure \ref{fig:pblender}) and \llm{} - MV(Avg.) (Figure \ref{fig:lblender}), show more balanced performance across all system types. Both methods demonstrate a near-uniform spread along the diagonal, indicating that the blending of multiple LLM outputs reduces biases in both overestimating and underestimating relevance.


\section{Conclusion and Future Work}
\label{sec:conclusion}

In this paper, we introduced \model, demonstrating how both variations -- \prompt and \llm -- are effective methods for evaluating LLM performance. By aggregating results from diverse prompts (\prompt) or multiple models (\llm), we reduce costs and improve correlation with human judgments. Our findings reveal that there is no single `best’ judge across all settings; however, both \prompt and \llm consistently perform well. Although our investigation was limited to a small number of evaluator settings, panel compositions, and prompts, we have shown that \model offers a robust alternative to relying on a single large model.

Due to resource and budget limitations, our experiments were restricted to a single dataset in this paper. Future work could extend \model by experimenting with a wider variety of prompts, datasets, and LLMs. Optimising panel selection and prompt strategies to balance quality and cost remains an open area for further research. Additionally, evaluating more advanced aggregation strategies -- such as leveraging other LLMs to make the final decision -- presents significant potential for future exploration.

\section*{Acknowledgements}
This work is supported by the Engineering and Physical Sciences Research Council [EP/S021566/1], the EPSRC Fellowship titled ``Task Based Information Retrieval'' [EP/P024289/1], and the Turing Fellowship scheme.

%
%
%
\bibliographystyle{splncs04}
\bibliography{references}
\end{document}